# From scale-dependent feedbacks to long-range competition alone: a short review on pattern-forming mechanisms in arid ecosystems.

Ricardo Martínez-García[1*], Cristóbal López[2]

[1]Department of Ecology and Evolutionary Biology, Princeton University, Princeton NJ 08544, USA

[2]IFISC, Instituto de Física Interdisciplinar y Sistemas Complejos, (CSIC-UIB) E-07122 Palma de Mallorca, Spain.

*Correspondence: ricardom@princeton.edu

**Abstract.**

Vegetation patterns are abundant in arid and semiarid ecosystems, but how they form remains unclear. One of the most extended theories lies in the existence of scale-dependent feedbacks (SDF) in plant-to-plant and plant-water interactions. Short distances are dominated by facilitative interactions, whereas competitive interactions dominate at larger scales. These feedbacks shape spatially inhomogeneous distributions of water that ultimately drive the emergence of patterns of vegetation. Even though the presence of facilitative and competitive interactions is clear, they are often hard to disentangle in the field, and therefore their relevance in vegetation pattern formation is still disputable. Here, we review the biological processes that have been proposed to explain pattern formation in arid ecosystems and how they have been implemented in mathematical models. We conclude by discussing the existence of similar structures in different biological and physical systems.

1. Introduction

Self-organization is ubiquitous in nature. Instances can be found at any spatiotemporal scale, from microbes to entire landscapes, and both in motile and sessile organisms (Solé & Bascompte 2006). One of the most deeply studied instances is vegetation patterning in arid ecosystems, in which order emerges at very large scales and under different biotic (vegetation species) and abiotic (soil types) conditions. Patterning in living systems is often a response to external stressors (Meron 2018) and the emergent structures therefore contain important information about the physical and biological processes occurring in the system. In this sense, changes in vegetation patterns have been theoretically proposed as an early-warning indicator for desertification and plant biomass loss (Scheffer & Carpenter 2003; Rietkerk et al. 2004), as well as a proxy for quantifying the ecosystems' response to environmental changes (Siteur et al. 2014).

Despite the large variety of regions in which vegetation patterns have been identified (see Borgogno et al. 2009; Deblauwe et al. 2008; Rietkerk & van de Koppel 2008 for a recompilation of pattern locations), the same shapes are consistently found across the globe; vegetation mainly forms spots interspersed with areas of bare soil as well as soil-vegetation labyrinthic patterns (Fig. 1). In the last years, an increasing number of theoretical studies have proposed a large variety of mechanisms to explain the emergence of patterns of vegetation. Most of these models use a continuum approach in which vegetation biomass is described as a continuous field that evolves in space and time. From the mathematical point of view, models can be grouped into two main classes: (i) Turing-like models that explicitly describe the water and vegetation dynamics via a pair of coupled partial differential equations, and (ii) kernel-based models that describe the dynamics of the vegetation using a single partial integro-differential equation (Borgogno et al. 2009). From a more biological perspective, the first class of models accounts for the existence of positive and negative feedbacks between water and vegetation, whereas in the second, those feedbacks are effectively incorporated in the existence of competitive and/or facilitative interactions among plants. More sophisticated models lying in each of these classes also include interactions between vegetation and other organisms (Bonachela et al. 2015; Tarnita et al. 2017; Pringle & Tarnita 2017), landscape topography (Klausmeier 1999; von Hardenberg et al. 2001), and different sources of environmental or demographic stochasticity (D'Odorico et al. 2006b; Martínez-García, Calabrese & López 2013; Butler & Goldenfeld 2009).

## 2. Turing-like models for water-vegetation dynamics

In 1952 Turing showed in his pioneering work on morphogenesis that an activation-inhibition interaction between two chemicals, coupled to differences in their diffusion coefficients, can lead to the formation of an inhomogeneous spatial distribution of the two substances (Turing 1952). In Turing's original model, the activator produces more of itself via an autocatalytic reaction and a second substance that inhibits the production of the activator and therefore balances its concentration (Fig. 2a). For the pattern to emerge, the inhibitor must diffuse faster than the activator, so that it inhibits the production of the activator over a long range and therefore confines the concentration of the activator locally (Fig. 2b). This activation-inhibition principle is thus scale-dependent; positive feedbacks dominate on short scales and negative feedbacks dominate on larger scales.

In our context, vegetation acts as the self-replicating activator and water as the inhibitor limiting resource. To discuss this family of models, we focus on the seminal work by Klausmeier 1999. Initially formulated to describe the formation of stripes of vegetation in sloping landscapes, it can be extended to flat grounds (Kealy & Wollkind 2012). The proposed pair of coupled partial differential equations is:

$$\frac{\partial W}{\partial t} = P - LW - RG(W)F(V)V + D_W \nabla^2 W, \tag{1a}$$

$$\frac{\partial V}{\partial t} = RJG(W)F(V)V - MV + D_V \nabla^2 V, \tag{1b}$$

where $W(r;t)$ and $V(r;t)$ represent soil water, respectively vegetation biomass; the spatial and temporal dependence in both fields has been omitted for simplicity in the notation. In Eq. (1a), water is continuously supplied at a precipitation rate $P$ and it is lost due to evaporation, at rate $L$, and to local uptake by plants. Water uptake is modeled by the term $RG(W)F(V)V$, in which $R$ is the plant absorption rate, $G(W)$ is the functional response of plants to water, and $F(V)$ is an increasing function that represents the positive feedback in water infiltration due to the presence of vegetation. Finally, water diffuses with a diffusion coefficient $D_W$. In Eq. (1b), vegetation biomass has a growth term that depends on the presence of water and a density-independent mortality term at rate $M$. $J$ is the yield of plant biomass per unit water consumed. In the original model, the simplest choices for the plant absorption rate and the response of plants to water are made: $G(W) = W$ and $F(V) = V$. Finally, the diffusion term with a diffusion coefficient $D_V$, represents plant dispersal.

A shared feature among all the models in this class, is the existence of a scale-dependent feedback acting similarly to Turing's activation-inhibition principle. Several positive water-vegetation feedbacks have been studied in the literature (Meron 2018; Meron 2016). For instance, in Eqs. (1) vegetation growth increases the infiltration of water through the function $F(V)$ and thus enhances the growth of more vegetation at a short scale. Overall, the effect of the positive feedbacks, regardless of the mechanism that they represent, is to enhance water availability in more vegetated areas. Negative feedbacks, however, represent an increased water consumption caused by vegetation growth, which inhibits further biomass growth. Since plant dispersal occurs over much shorter scales than water diffusion (HilleRisLambers et al. 2001; Rietkerk et al. 2002), the negative feedback occurs at a much larger spatial scale ($D_W \gg D_V$).

Due to the similarity between the SDF mechanism and Turing's activation-inhibition principle, vegetation-water models provide the full set of patterns characteristic of Turing's model: as the precipitation, which is the control parameter for the aridity of the ecosystem, decreases, vegetation biomass transitions from being homogeneously distributed to arrange leaving gaps of bare soil; then to labyrinthic patterns, and finally to form a matrix of spots that are interspersed with bare soil (von Hardenberg et al. 2001). In addition, and depending on the initial condition used to perform the numerical integration of the model, ring-like structures are also observed in the transient toward stationary spotted patterns (Meron et al. 2004).

### 3. Spatially nonlocal models: a kernel-based approach

Turing-like models, in which water dynamics is explicitly included, allow a direct identification of the model parameters with the processes that they represent. Alternatively, water-vegetation feedbacks can be implicitly described as plant-to-plant interactions. This leads to a new class of models that use a single partial differential equation to describe the spatiotemporal dynamics of the vegetation alone. Plant-to-plant interactions occur on a finite range and are represented via nonlocal (intergral) terms. Therefore, the dynamics of the vegetation at any point of the space depends on the presence of vegetation at other positions. The properties of this coupling, such as whether it enhances or inhibits plant growth as well as its spatial range, are contained in a kernel function, denoted by *G*.

Kernel-based models allow for a more straightforward assessment of the mechanisms that mediate plant-to-plant interactions and their role in pattern formation via different choices for the kernel function. They can be classified depending on how the kernel is introduced in the equation and the mechanisms it accounts for.

### 3.1. Kernel-based models with competitive and facilitative interactions

Using trees to illustrate scale-dependent kernels, the facilitation range is usually assumed to be determined the crown radius, while the competition range is related to the lateral root length (Fig. 3a). The kernel is often defined as the addition of two Gaussian functions with different widths, with the wider being inverted to account for the longer range of competitive interactions (D'Odorico et al. 2006a) (Fig. 3b). Given the analogy between these kernels and the ones used to model neural processes, including stripe formation in the visual cortex, these models are also termed *neural models* (Murray 2002).

Models in this family can be classified depending on whether the spatial coupling (non-local interactions) enters in the equations linearly (D'Odorico et al. 2006a) or multiplicatively (Lefever & Lejeune 1997). In the simplest linear case, the spatial coupling is added to the local dynamics,

$$\frac{\partial V}{\partial t} = h(V) + \int_\Omega G(r',r)[V(r') - V_0], \qquad (3)$$

the first term in the right-hand side of Eq. (3) describes the local dynamics of the vegetation, i.e., the dynamics of *V* at a given position and independently of the amount of vegetation surrounding it, whereas the second term describes the spatial coupling, i.e., the interactions between vegetation at a position *r* and the rest of the system, as denoted by the integral over the whole space *Ω*. In the absence of the spatial coupling, vegetation density increases or decreases at each point of the space depending on the sign of *h(V)*. Equivalently, the spatial coupling may have a positive or negative effect on vegetation growth depending on its sign, which is determined both by the sign of the kernel function *G* (Fig. 3b) and the difference between the vegetation density at a given position, *V(r')*, and $V_0$. The shape of the kernel function *G* is thus responsible of the growth or decay of inhomogeneities in the spatial distribution of vegetation.

Assuming kernels like the one in Fig. 3b (positive close to the focal plant and negative far from it), perturbations in the vegetation density around $V_0$ are locally enhanced if they are larger than $V_0$ and attenuated otherwise. As a result, the homogeneous state losses its stability and spatial inhomogeneities arise in the system. Long-range inhibitory interactions, together with nonlinear terms in the local term *h(V)* avoid the indefinite growth of the perturbations and stabilize the pattern (Fig. 3c). Finally, even though neural models impose an upper bound to the vegetation density, they allow negative values in *V*, which are biologically nonsensical. To avoid this issue, numerical integrations of Eq. (3) always include and artificial bound at $V = 0$ such that vegetation density is reset to zero whenever it becomes negative.

As an alternative to artificially bounding the domain of the vegetation density, modulating the spatial coupling with nonlinear terms avoids negative values for the vegetation density. The pioneering model developed in (Lefever & Lejeune 1997) instantiates this approach. It describes the growth-death spatiotemporal dynamics of a single vegetal species,

$$\frac{\partial V}{\partial t} = F_1(V)F_2(V) - F_3(V), \qquad (4)$$

where $F_1$, $F_2$ and $F_3$ describe vegetation growth, plant-to-plant and plant-by-environment inhibitory interactions, and vegetation loss respectively. Each of these three functions are modulated by integral terms with different kernel functions that act at different scales to account for the non-local nature of each process. Since the authors set the scale of the inhibitory interactions to be larger than that of the positive interactions, the model includes a SDF with short-range facilitation and long-range competition. Moreover, since each function *F* is modulated by the local density of vegetation, the variable *V* has a natural lower bound and cannot take negative values. Expanding upon this work, several other models have introduced non-linear spatial couplings *via* integral terms (Ruiz-Reynés et al. 2017; Fernandez-Oto et al. 2014; Couteron et al. 2014; Escaff et al. 2015), even some have combined a Turing-like approach with non-local interactions (Gilad et al. 2004).

### 3.2. Kernel-based models with purely competitive interactions

In previous sections, we invoked the existence of SDFs in the interactions among plants to explain pattern formation. However, competition and facilitation usually act simultaneously and are hard to disentangle (Veblen 2008; Barbier et al. 2008). Moreover, some studies have highlighted the importance of long-range negative feedbacks on pattern formation (Rietkerk & van de Koppel 2008; Koppel et al. 2006), suggesting that short-range positive feedbacks might be secondary actors that mostly increase the sharpening of the clusters (Eppinga et al. 2009). Following these arguments, a family of purely competitive models was proposed (Martínez-García et al. 2014; Martínez-García, et al. 2013), aiming to unveil the minimal set of processes that could drive the emergence of patterns of vegetation in arid and semiarid ecosystems.

#### 3.2.1  Nonlocal linear spatial coupling

Inspired by the kernel-based models with short-range facilitation and long-range competition discussed in the previous sections, the simplest formulation of purely competitive models also accounts for a linear spatial coupling. Models in this family can be written as

$$\frac{\partial V}{\partial t} = D\nabla^2 V + \beta V\left(1 - \frac{V}{V_{max}}\right) + \lambda \int G(r,r')V(r',t)dr' \tag{5}$$

where the first term on the right side represents long-range seed dispersal. The second one is a logistic-like growth term in which the growth-limiting factor $(1-V/V_{max})$ accounts for local seed dispersal and thus represents local competition for space; $\beta$ is the seed production rate. The third term is the spatial coupling. Since the kernel function $G(r,r')$ only represents competitive interactions, it needs to be a negative function; $\lambda$ is a positive parameter that controls the intensity of the competition. Due to the linear non-local term, $V$ can take negative values and an artificial bound at $V = 0$ has to be imposed for numerical integrations of Eq. (5). Since $\lambda$ and $V$ are always positive and $G$ negative, the spatial coupling is always negative and therefore represents a contribution to biomass loss. Assuming isotropic systems, a typical choice for the kernel $G$ is a negative box-like function of the distance coordinate, $|r - r'|$, as shown in Fig. 3c. However, using a linear stability analysis of Eq. (5), it can be shown that patterns may form for many other kernels. In addition, the shape of the patterns, either labyrinths or spots of vegetation, resembles those obtained in Turing-like models (Fig. 4a, b).

### 3.2.1. Nonlocal Nonlinear spatial interactions

Alternatively, one can introduce the non-local interactions in a nonlinear fashion, either modulating biomass growth or loss. In both cases we recover the same sequence of patterns. We first discuss the model with a nonlocal birth term introduced in (Martínez-García et al. 2013), which assumes that population growth follows a sequence of seed production, local dispersal and establishment, and that population declines at a constant rate,

$$\frac{\partial V}{\partial t} = P_E(\tilde{V},\delta)\beta V(1 - V) - \alpha V, \tag{6}$$

where $\beta$ is the seed production rate, $\delta$ is the competition-strength parameter and $\tilde{V}(r,t)$ is the average density of vegetation around the focal position $r$, termed nonlocal vegetation density in the following. It is calculated as

$$\tilde{V}(r,t) = \int g(r,r')V(r,t)dr'. \tag{7}$$

It is important to remark here the difference between the kernel function $g$ in Eq. (7) and previously defined kernel functions $G$. In both cases, they are called kernels function because the enter in the integral part of the equation, but they represent different magnitudes. $G$ contains information about plant-to-plant interactions as a function of the distance between them; $G$ positive represents facilitation and $G$ negative, competition (Fig. 3b). In contrast, $g$ only defines an area of influence of a focal plant, typically determined by the characteristic scale of the function, $R$, and how this influence changes on space. Therefore, $g$ is always a positive function normalized to one, regardless of the nature of the interactions considered in the model. Finally, we will limit here to isotropic cases in which $g(r, r') = g(|r - r'|)$.

While vegetation loss is assumed to occur at a constant rate $\alpha$, population growth is modelled through a sequence of seed production, local dispersal, and establishment. Mathematically, this is represented by the three factors that contribute to the first term in the right side of Eq. (6). Initially, plants produce seeds at a constant rate $\beta$; if one assumed that every seed establishes and gives rise a to a new plant, then biomass growth would be represented by a $\beta V$ term alone. However, the model considers two competitive mechanisms that act after seed production. First, local seed dispersal and competition for space. Following the rationale of Eq. (5), we assume that the availability of space limits the maximum density at each point of the space to a maximum value $V_{max}$. For simplicity, this maximum value can be rescaled such that $V_{max} = 1$ and the proportion of available space at a position $r$ is $1-V(r,t)$. This explains the $1-V(r,t)$ growth-limiting factor in Eq. (6). Second, competition for resources through a plant establishment probability. The model assumes that once space limitations have been overcome, seeds need to get over the competition for resources, mostly water, with already established plants. This process is introduced *via* a probability of establishment term, $P_E$. Since water intake is mediated by the roots, $P_E$ is a function of the density of vegetation modulated by the competition-strength parameter, $\delta$, which essentially represents the limitation of resources. This means water abundant conditions ($\delta = 0$), competition for water is not intense and $P_E = 1$ whereas new plants cannot establish in the limit of extremely arid landscapes ($\delta = \infty$). Furthermore, more crowded neighborhoods also represent more competitive scenarios and the probability of establishment thus decreases with increasing vegetation density,

$$\left(\frac{\partial P_E}{\partial \tilde{V}}\right) < 0. \tag{8}$$

A complete description of the model needs to specify the kernel function $g$ and the probability of establishment $P_E$. However, from this general formulation it is possible to proof the existence of patterns provided that the influence of plants on the competition for resources (i.e., the form of the function $g$) meets a series of conditions (Martínez-García, Calabrese, Hernández-García, et al. 2013). Specifically, a necessary condition is that the Fourier transform of $g$ becomes negative for some wavenumber, which is true for any function that presents a discontinuity at a distance from the focal plant. This distance would be related to the typical root length. An instantiation of these kernels is a top-hat function (inverted Fig. 3c). Once the kernel function $g$ meets this condition, the parameterization of the model determines whether patterns form or not. For low values of the competition strength $\delta$, a homogeneous state with $V \neq 0$ is stable; as $\delta$ increases, the homogeneous state becomes unstable and the stationary distribution of vegetation consists of a pattern of stripes of vegetation interspersed with stripes of bare soil. If $\delta$ continues to increase, the spatial pattern changes to spots of vegetation scattered on a background of bare soil. These spots arrange forming a hexagonal lattice (Fig. 4b, d) similar to those reported in several territorial animal species (Pringle & Tarnita 2017). Finally, in the limit of very strong competition, the only stable state is a desert state with $V = 0$.

An alternative formulation in this family of models is to assume that the competition for resources influences the probability of a plant to die instead of the plant establishment probability. Mathematically, this means that $P_E = 1$ and the death term is modulated by a death probability $P_D$,

$$\frac{\partial V}{\partial t} = \beta V(1 - V) - \alpha P_D(\tilde{V}, \delta)V \tag{9}$$

Since the role of $P_D$ is to enhance plant mortality instead of to inhibit its growth, its properties are broadly the opposite to those imposed to $P_E$ (Martínez-García et al. 2014). Death terms modulated by non-local competition have been previously shown to favor clustering of individuals in population models (Birch & Young 2006).

In this section, we have discussed two different implementations of nonlocal interactions that result in the same sequence of patterns. The conditions needed to have patterns are entirely encapsulated in the shape of the spatial interactions through the negativity of the Fourier transform of the kernel function $g$. Moreover, a spectral analysis of the patterns indicates that they have a periodicity between one and two times the range of the spatial interactions, $R$. These two results suggest that the symmetry-breaking instability of the homogeneous state and the transition to patterns is encoded in the non-local term, rather than in some sort of nonlinearity in the local dynamics of the model. For certain choices of the kernel function $g$, inhomogeneities in the distribution of vegetation are enhanced through the formation of *exclusion areas*: regions of the space in which the density of roots and therefore plant-to-plant competition is extremely high (Martínez-García et al. 2014; Pigolotti et al. 2007). The formation of these areas, driven by competition alone and without any facilitative interaction, also provides an argument for the distance between clusters. A random and spatially heterogeneous distribution of vegetation will have local maxima, representing regions of the space with the highest density of plants. If two of these maxima are separated a distance larger than $R$ but smaller than $2R$, then plants within one cluster do not interact with plants in the other cluster. This is because the distance between clusters is larger than the interaction range $R$ (twice the typical length of the roots) (Fig. 3e). However, at the halfway between both clusters, there is a region in which germinating seeds compete with plants in both clusters in order to establish a new plant. Similarly, when non-local competition is implemented in the death term, competition is stronger in the exclusion areas and plant biomass tends to disappear from these regions. Once individuals disappear from the region between patches, plants in the cluster experience a weaker competition for resources, which results in a positive feedback that increases the biomass inside the patch.

## 4. Conclusions

Despite the differences among the models reviewed here and the biological ingredients they consider, all the models recover the same spectrum of patterns, which highlights the model and mechanism-independence of the patterns. In this context, two different lines of research emerge. On the one hand, biologically grounded studies should aim to combine system-specific models with empirical measures of vegetation-mediated feedbacks. On the other hand, theoretical efforts should move toward reconciling Turing-like and kernel-based models and establishing a direct connection between the mechanisms included in each of them. Such a relationship is still lacking, except for certain approximations of the neural-models, in which the nonlocal term is expanded in a series of differential operators (Borgogno et al. 2009). To our knowledge, any attempt to derive a kernel-based model for the vegetation field starting from a more fundamental description that considers water and vegetation dynamics has been unsuccessful in reproducing the appropriate shape of the kernels (Martínez-García et al. 2014).

From a mechanistic point of view, besides being successful in recapitulating the variety of patterns of vegetation observed in nature, SDF and purely competitive models have been reported as drivers of spatial self-organization in many other systems. A combination of *attractive* and *repulsive* forces acting on different scales is, for instance, responsible of the formation of regular stripes in mussel beds. High mussel densities increase the competition for nutrients over long distances but they facilitate mussel-attachment to the sediment on the shorter range (van de Koppel et al. 2005; Rietkerk & van de Koppel 2008). Other models to study the formation of different structures in animal grouping also rely on similar attraction-repulsion principles (Couzin et al. 2002; Couzin 2003; Martínez-García et al. 2015; Liu et al. 2013; Liu et al. 2016; Vicsek & Zafeiris 2012). On the other hand, even though the fact that only competitive interactions may lead to clustering and pattern formation seems counterintuitive, it has been observed in several scenarios as well. Among biological systems, clustering in the niche space has been predicted in the context of species competing for shared resources (Scheffer & van Nes 2006; Pigolotti et al. 2007). Other examples come from the physical sciences, such as the formation of the so-called cluster crystals in some molecules and colloids that interact via effective repulsive forces (Mladek et al. 2006; Likos et al. 2007; Klein et al. 1994; Delfau et al. 2016). Patterning in these disparate systems shares common properties: competition induces a hexagonal distribution of the clusters and the transition to patterns is mathematically controlled by the sign of the Fourier transform of the kernel function. Deepening on the generality of these properties for repulsion-induced clustering arises as a challenging line for future research.

Overall, this compendium of systems shows that seemingly identical patterns can emerge in different scenarios and from different interactions. This is especially important from an ecological point of view, since patterns that seem identical but originate from different mechanisms could have completely different (eco)system-level consequences that would require completely different managing strategies.

**Acknowledgements**

We greatly acknowledge our close collaborators Justin M. Calabrese and Emilio Hernández-García. We are also thankful to Corina E. Tarnita, Federico Vazquez, Damia Gomila, and Miguel Angel Muñoz, for insights and collaborations on these and related topics. RMG is a Life Sciences Research Foundation postdoctoral fellow. This work is funded by the Gordon & Betty Moore Foundation through grant GBMF2550.06 to RMG and Ministerio de Economía y Competitividad and Fondo Europeo de Desarrollo Regional through project CTM2015-66407-P (MINECO/FEDER) to CL.

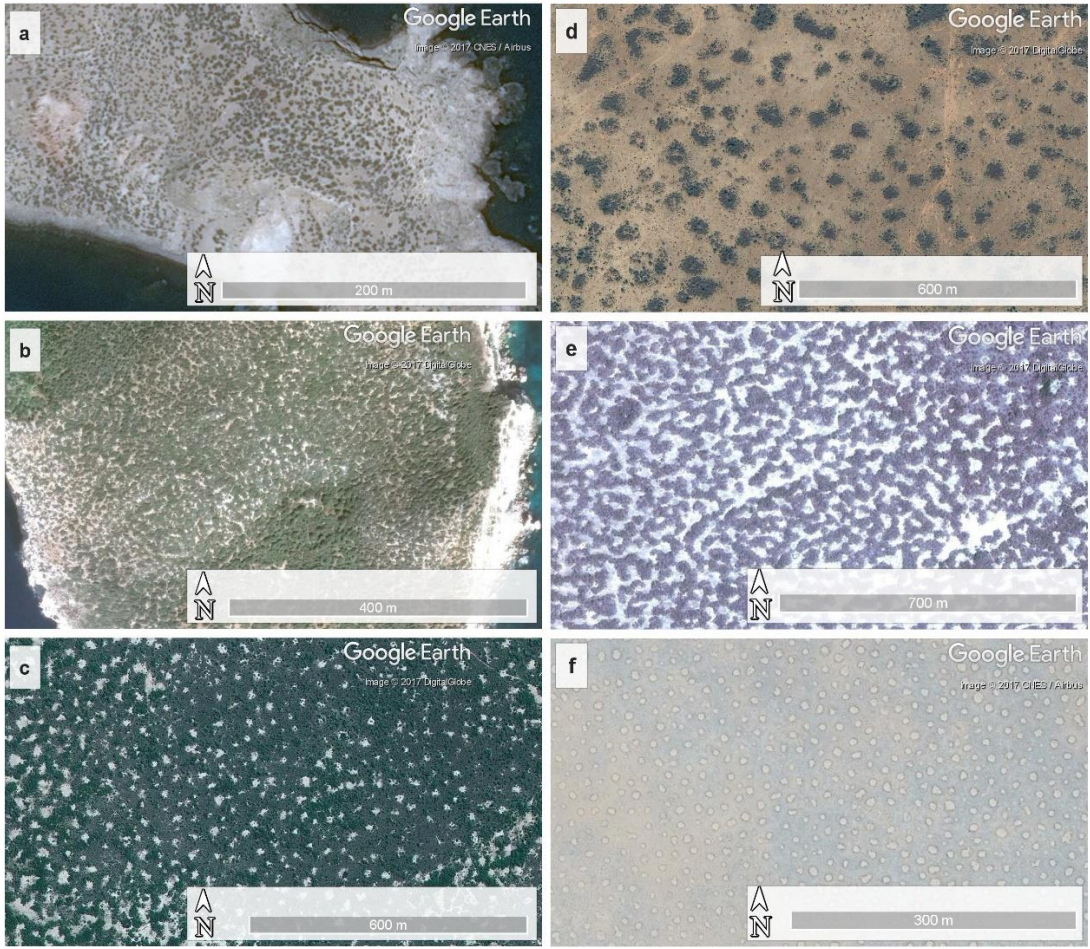

**Figure 1. Aerial imagery of representative vegetation patterns.** Imagery: Google, DigitalGlobe and CNES Airbus. a) Spot pattern in the Chafarinas Archipielago (Spain); 35°10'44.73"N, 2°26'26.54"W. b) Labyrinth pattern in Cabrera Archipielago (Spain); 39°10'45.87"N, 2°57'55.73"E. c) Gap pattern in the Republic of Niger; 13°11'29"N, 1°15'9.07"E, d) spot pattern in Chad 11°52'9.52" N, 15°59'42.7"E, e) labyrinth pattern in the Republic of Niger; 13°6'8.29"N, 213'19.12"E (Bailey 2011), f) fairy circles (gap) in the Namibian desert; 24°57'S 15°55'E

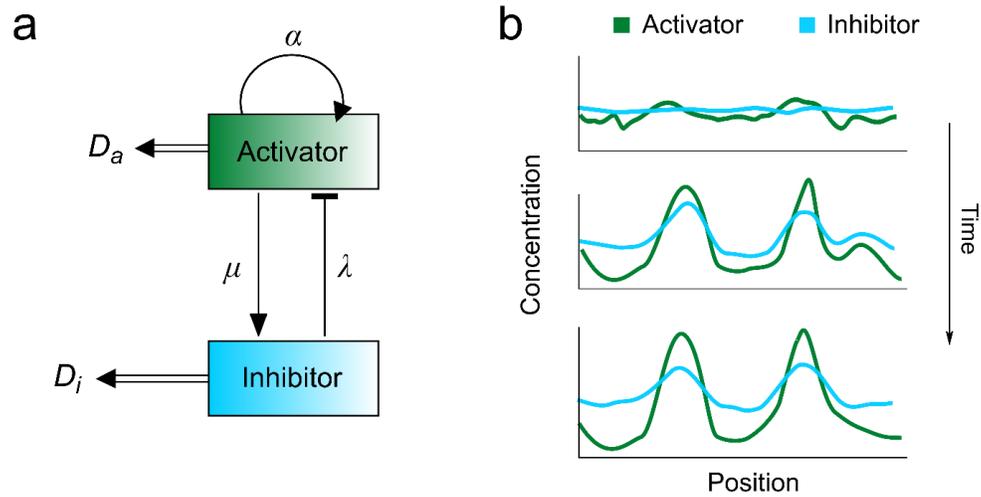

**Figure 2. Turing principle. a**) Schematic of the Turing activation-inhibition principle. The activator, with diffusion coefficient $D_a$, produces the inhibitor at rate $\mu$ as well as more of itself at rate $\alpha$ *via* an autocatalytic reaction. The inhibitor degrades the activator at rate $\lambda$ and diffuses at rate $D_i > D_a$. **b**) Schematic of the pattern-forming process in a one-dimensional system.

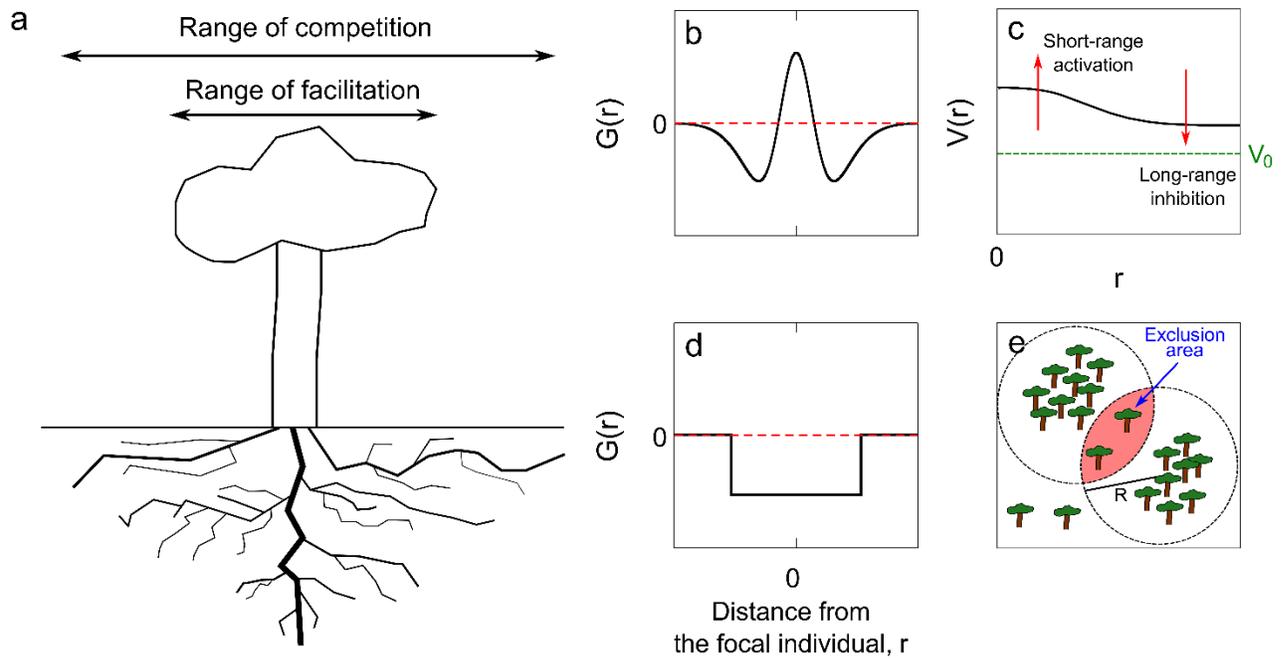

**Figure 3. Kernel-based model properties. a**) Characteristic range of facilitation and competition (Borgogno et al. 2009). **b**) Short-range facilitation and long-range competition kernel. **c**) Symmetry-breaking instability mechanism in models with facilitative and competitive interactions. **d**) Purely competitive kernel. **e**) Schematic of the formation of exclusion areas.

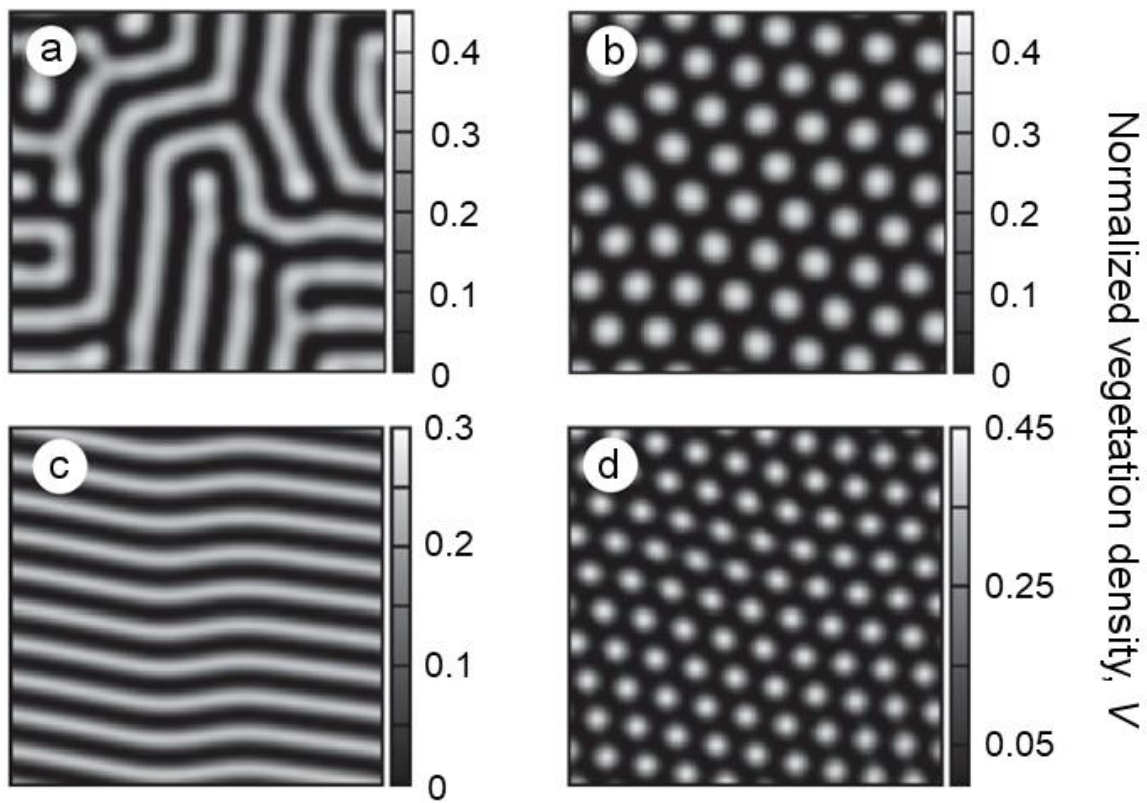

**Figure 4. Patterns generated by kernel-based models with purely competitive interactions in the death term.** Non-local interactions enter linearly (a, b) or non-linearly (c, d) in the model. Simulation details and parameterization as in (Martínez-García et al. 2014).